\documentclass[a4paper,12pt]{article}
\usepackage[pctex32]{graphics}

\begin{document}
\topmargin 0pt \oddsidemargin 0mm
\newcommand{\beq}{\begin{equation}}
\newcommand{\eeq}{\end{equation}}
\newcommand{\beqa}{\begin{eqnarray}}
\newcommand{\eeqa}{\end{eqnarray}}
\newcommand{\sr}{\sqrt}
\newcommand{\fr}{\frac}
\newcommand{\mn}{\mu \nu}
\newcommand{\G}{\Gamma}

\begin{titlepage}

\vspace{5mm}
\begin{center}
{\Large \bf  Phase transition for black holes with scalar hair and
topological black holes} \vspace{12mm}

{\large   Yun Soo Myung \footnote{e-mail
 address: ysmyung@inje.ac.kr}}
 \\
\vspace{10mm} {\em Institute of Mathematical Science and School of
Computer Aided Science \\ Inje University, Gimhae 621-749, Korea}
\end{center}
\vspace{5mm} \centerline{{\bf{Abstract}}}
 \vspace{5mm}
We study phase transitions between  black holes with scalar hair
and topological black holes in asymptotically anti-de Sitter
spacetimes. As the ground state solutions, we introduce the
non-rotating BTZ black hole in three dimensions and topological
black hole with hyperbolic horizon in four dimensions. For the
temperature matching only, we show that the phase transition
between black hole with scalar hair (Martinez-Troncoso-Zanelli
black hole) and topological black hole is  second-order by using
differences between two free energies. However, we do not identify
what order of the phase transition between scalar and non-rotating
BTZ black holes occurs in three dimensions, although there exists
a possible decay of scalar black hole to non-rotating BTZ black
hole.
\end{titlepage}
\newpage
\renewcommand{\thefootnote}{\arabic{footnote}}
\setcounter{footnote}{0} \setcounter{page}{2}

\section{Introduction}
Topological black holes in asymptotically anti-de Sitter
spacetimes were first found in three and four
dimensions~\cite{Lemos}. Their black hole horizons are Einstein
spaces of spherical ($k=1$), hyperbolic ($k=-1$), and flat ($k=0$)
curvature for higher dimensions more than three~\cite{Vanzo,BLP}.
The standard equilibrium and off-equilibrium thermodynamic
analysis is possible to show that they are treated as the extended
thermodynamic system  even though their horizons are not
spherical.

 The Schwarzschild black hole with spherical horizon
  is  in an unstable equilibrium with the  heat
reservoir of the temperature $T$~\cite{GPY}. Its fate under small
fluctuations will be either to decay to  hot flat space by Hawking
radiation or to grow without limit by absorbing thermal radiation
in the heat reservoir~\cite{York}. This means that an isolated
black hole is never in thermal equilibrium.  There exists a way to
achieve a stable black hole in an equilibrium with the heat
reservoir. A black hole could be rendered thermodynamically stable
by placing it in four-dimensional anti-de Sitter (AdS$_4$)
spacetimes. An important point to understand  is how a stable
black hole with positive specific heat could emerge from thermal
radiation through a phase transition. This is known as  the
Hawking-Page phase transition between thermal AdS space and
Schwarzschild-AdS black hole~\cite{HP,BCM,Witt}, which is a
typical example of the first-order phase transition.

It was proposed that a phase transition between AdS black hole
with hyperbolic horizon  and AdS massless black hole with
degenerate horizon is unlikely to occur~\cite{myungsds}. Moreover,
the phase transition between AdS black hole with Ricci-flat
horizon and AdS soliton has been proposed for a candidate of phase
transition in topological black holes~\cite{HM,SSW}.

Recently,  a black hole solution  with a minimally coupled
self-interacting scalar was found in   AdS$_4$
spacetimes~\cite{MTZ}. This is called the
Martinez-Troncoso-Zanelli black hole (MTZ). Its horizon geometry
is a surface of hyperbolically constant curvature. This solution
dressed by a scalar reminds us of the topological black hole  with
hyperbolic horizon (TBH). It was conjectured that there is a
second-order phase transition between MTZ and
TBH~\cite{KMPS,SWSLC}. Furthermore, it was argued that quasinormal
modes show a signal to this  phase transition.

On the other hand, there exists a black hole solution to
three-dimensional gravity with a minimally coupled
self-interacting scalar, called the scalar black hole
(SBH)~\cite{HMTZ,GMT}. It was observed that the SBH can decay into
the non-rotating Banados-Teitelboim-Zanelli black hole
(NBTZ)~\cite{BTZ}. We note that in three dimensions, there is no
distinction between AdS black hole with spherical horizon (NBTZ)
and those with hyperbolic   and  flat horizons~\cite{SSW}. Hence
we expect to have  a phase transition  between SBH and NBTZ.

In this Letter,  we show that the phase transition between MTZ and
TBH is  second-order by using temperature matching and off-shell
free energies. However, we do not identify what order of the phase
transition between SBH and NBTZ occurs in three dimensions,
although there exists a possible decay of SBH into NBTZ.

Our study is based on the on-shell observations of temperature,
mass, heat capacity and  free energy as well as the off-shell
observations of  generalized (off-shell)  free energy.  In
general, the on-shell thermodynamics implies equilibrium
thermodynamics and thus the first-law of thermodynamics holds for
this case. Hence it describes relationships among thermal
equilibria, but not the transitions between equilibria. On the
other hand, the off-shell thermodynamics is designed for the
description of off-equilibrium configurations~\cite{FFZ,off}.  We
note that the first-law of thermodynamics does not hold for
off-shell thermodynamics. This approach is suitable for the
description of phase transitions between thermal equilibria.
Especially, the off-shell free energy shows the phase transition
characteristics more manifestly than on-shell free energy.
Introducing this quantity leads to that the temperature dependence
of phase transition is clearly understood. In this work, we focus
on how the second-order phase transition do occur between  MTZ and
TBH.

The organization of this work is as follows. In section 2, we
first review thermodynamics of SBH and NBTZ in three-dimensional
AdS spacetimes. We discuss the phase transition between them by
introducing off-shell free energies. In section 3, we review
thermodynamics of MTZ and TBH in four-dimensional AdS spacetimes
by introducing coordinate and temperature matchings. Then we study
the phase transition between them by considering the difference of
their off-shell free energies. We discuss a possible connection
between quasinormal modes and phase transitions in section 4.
Finally, we summarize our results in section 5.

\section{Transition between SBH and  NBTZ }

We wish  to study  the scalar black hole  in three-dimensional AdS
(AdS$_3$) spacetimes. For this purpose,  we introduce the action
in three-dimensional spacetimes~\cite{HMTZ} \beq
I_3[g,\phi]=\frac{1}{\pi G_3} \int
d^3x\sqrt{-g}\Bigg[\frac{R}{16}-\frac{1}{2}(\nabla\phi)^2-V_\nu(\phi)\Bigg]
\label{SDS} \eeq where the potential $V_\nu(\phi)$ is given by
\beq V_\nu(\phi)=-\frac{1}{8l^2}\Big( \cosh^6\phi+\nu \sinh^6
\phi\Big).\eeq Here $\nu$ is a parameter and $l$ is the curvature
radius of AdS$_3$ spacetimes. In the case of $\phi=0$, we have the
BTZ black hole solution. For $\nu \ge -1$, there is a circularly
symmetric black hole solution dressed with the scalar field
\beq\bar{\phi}(r)=\tanh^{-1}\sqrt{\frac{B}{H(r)+B}}.\eeq
 Here
$B$ is a non-negative constant and $H(r)=(r+\sqrt{r^2+4Br})/2$.
Then the line element of the scalar black hole is give by \beq
ds^2_{S}=-\Bigg(\frac{H}{H+B}\Bigg)^2F(r)dt^2+
\Bigg(\frac{H+B}{H+2B}\Bigg)^2\frac{dr^2}{F(r)}+r^2d\varphi^2\eeq
with the metric function \beq \label{smetric}
F(r)=\frac{H^2}{l^2}-\frac{1+\nu}{l^2}\Bigg(3B^2+\frac{2B^3}{H}\Bigg).
\eeq The causal structure of this geometry is the same as for the
NBTZ~\cite{BTZ}. The event horizon is located at \beq \label{3EH}
r_{+}= B \theta_\nu \ge 0 \eeq where $\theta_\nu$ is the
first-zero of the Schuster function of order $\nu$ defined by \beq
\theta_\nu=2\Big(z\bar{z}\Big)^{2/3}\frac{z^{2/3}-\bar{z}^{2/3}}{z-\bar{z}}\eeq
with $z=1+i\sqrt{\nu}$ and its complex conjugate $\bar{z}$. As is
shown in Fig. 1, $\theta_\nu$ is a monotonically increasing
function of $\nu$ and  grows as $\sqrt{\nu}$ asymptotically.
Importantly, we observe that $\theta_{-1}=0$ and $\theta_{0}=4/3$.
From Eq.(\ref{smetric}), we have $F(r)= \frac{r^2}{l^2}\to 0$ as
$B\to 0(H(r)\to r)$, where the massless degenerate black hole is
found.
\begin{figure}[t!]
 \centering
   \includegraphics{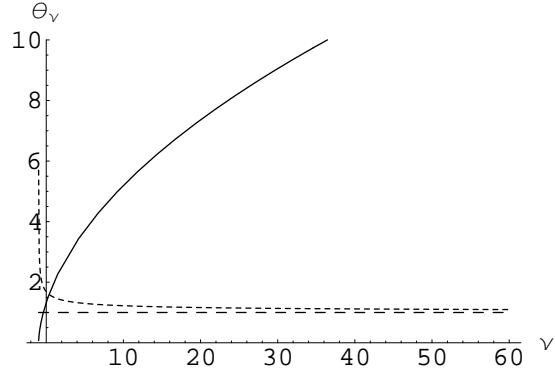}
    \caption{Plot of $\theta_\nu$ (solid curve) and $ \frac{3(1+\nu)}{\theta_\nu^2}$ (small dashed curve).
    The large dashed line means $1$, as the asymptotic line for $ \frac{3(1+\nu)}{\theta_\nu^2}$.  } \label{fig1}
\end{figure}

 Thermodynamic quantities of the SBH are Hawking temperature $T_S$, mass $M_s$, heat capacity $C_S$, entropy $S_S$,
 on-shell free energy $F_S$, and off-shell free energy $F^{off}_S$
 \beqa \label{SBH}T_S&=&\frac{r_+}{2\pi
l^2}\frac{3(1+\nu)}{\theta_\nu^2},~M_S=\frac{r_+^2}{8G_3
l^2}\frac{3(1+\nu)}{\theta_\nu^2},~C_S=\frac{\pi r_+}{2G_3}=S_S,\\
F_S(r_+)&=&M_S-T_S S_S=-\frac{r_+^2}{8G_3
l^2}\frac{3(1+\nu)}{\theta_\nu^2},~F^{off}_S(r_+,T)=M_S-T S_S.
\eeqa We note that the first-law of thermodynamics $dM_S=T_SdS_S$
holds for the SBH. Thanks to this, we check that for
$dF_S^{off}=0$ (saddle point), one has thermal equilibrium point
of $T=T_S$ and thus  $F_S^{off}(r_+,T)=F_S(r_+)$. Here we derive
the free energy as a function of $T_S$ \beq F_S(T_S)=- \frac{(2
\pi l^2 T_S)^2}{8G_3}\frac{\theta_\nu^2}{3(1+\nu)}. \eeq We
observe an inequality of $\frac{3(1+\nu)}{\theta_\nu^2}>1$ for
$\nu \ge -1$ from Fig. 1.

On the other hand, for $\phi=0$, we have the NBTZ whose
thermodynamic quantities as Hawking temperature $T_B$, mass $M_B$,
heat capacity $C_B$, entropy $S_B$,
 on-shell free energy $F_B$, and off-shell free energy $F^{off}_B$
 \beqa \label{NBTZ}T_B&=&\frac{\rho_+}{2\pi
l^2},~M_B=\frac{\rho_+^2}{8G_3
l^2},~C_B=\frac{\pi \rho_+}{2G_3}=S_B,\\
F_B(\rho_+)&=&M_B-T_B S_B=-\frac{\rho_+^2}{8G_3
l^2},~F^{off}_B(\rho_+,T)=M_B-T S_B \eeqa with $\rho_+ \ge 0 $. It
is obvious that the first-law holds for the NBTZ. Also we find the
free energy as a function of $T_B$ \beq ~F_B(T_B)=-\frac{(2 \pi
l^2 T_B)^2}{8G_3}.\eeq The coordinate matching of $r_+=\rho_+$
implies  the heat capacity (entropy) matching of
$C_S=C_B(S_S=S_B)$. Choosing $G_3=l/4$ with $l=1$ and $
\frac{3(1+\nu)}{\theta_\nu^2}=1.1$ for $\nu=56.12$, the behaviors
of thermodynamic quantities  are shown in Fig. 2. We note that
there is an inconsistency between $F_B(r_+)\ge F_S(r_+)$ and
$F_B(T_B)\le F_S(T_S)$. Furthermore,  we evaluate  differences
between  free energies \beqa \Delta
F(r_+)&=&F_B(r_+)-F_S(r_+)=-\frac{r_+^2}{8G_3
l^2}\Big[1-\frac{3(1+\nu)}{\theta_\nu^2}\Big],\\
 \Delta F^{off}(r_+,T)&=&F^{off}_B(r_+,T)-F^{off}_S(r_+,T)=-\Delta
F(r_+),\eeqa where the latter is independent of the external
temperature $T$. Thus we cannot describe  any phase transition
between them under the coordinate matching.

\begin{figure}[t!]
 \centering
   \includegraphics{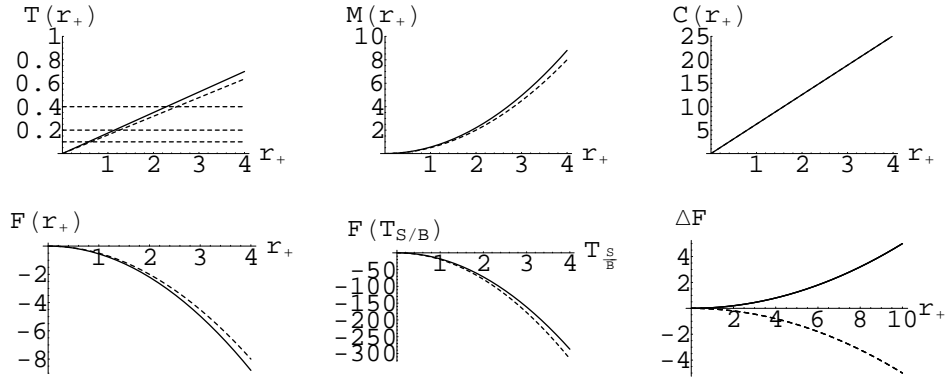}
    \caption{Coordinate matching of $\rho_+=r_+$. Hawking temperature $T(r_+)$ with external temperatures $T=0.4,0.2,0.1$, mass $M(r_+)$,  heat capacity $C(r_+)$, free energy
    $F(r_+)$ and $F(T_{S/B})$  for SBH (solid curves) and
    NBTZ (dashed curves).
    The right graph of  the bottom panel shows the difference of the on-shell (solid curve)  free
    energies  $\Delta F(r_+)$ and that of off-shell (dashed curve) free
energies of $\Delta F^{off}(r_+,T)$.
     } \label{fig2}
\end{figure}
In order to resolve this situation, we need another matching for
temperatures: $T_S=T_B$, which means that $\rho_+=r_+
\frac{3(1+\nu)}{\theta_\nu^2}$ with $\rho_+ \ge r_+$.  Then we
find  consistent inequalities for free energy \beq
\label{ief}F_B(r_+)\le F_S(r_+) ~~{\rm and}~~ F_B(T_B)\le
F_S(T_B).\eeq
\begin{figure}[t!]
 \centering
   \includegraphics{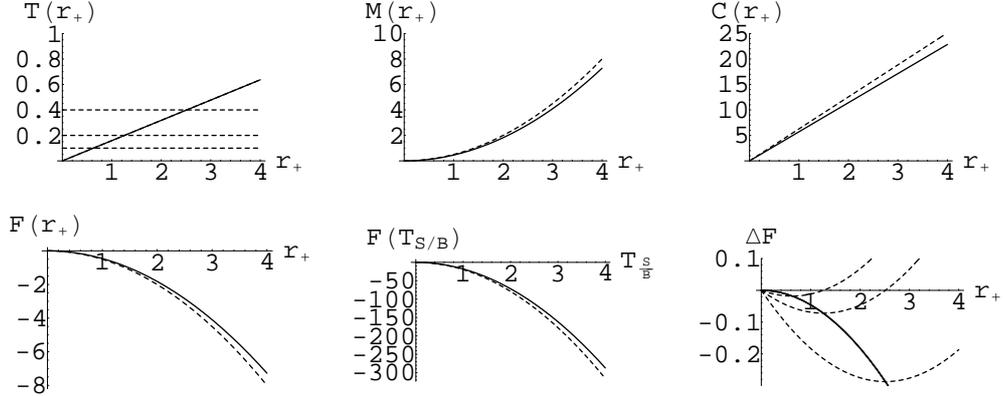}
    \caption{Temperature matching $\rho_+=r_+\frac{3(1+\nu)}{\theta_\nu^2}$.
     Hawking temperature $T(r_+)$ with external temperatures $T=0.4,0.2,0.1$, mass $M(r_+)$,  heat capacity $C(r_+)$, free energy
    $F(r_+)$ and $F(T_{S/B})$  for SBH (solid curves) and
    NBTZ (dashed  curves).
   The right graph of  the bottom panel shows the difference of the on-shell (solid curve)  free
    energies  $\Delta F(r_+)$ and that of off-shell (dashed curves) free
energies of $\Delta F^{off}(r_+,T)$ with $T=0.1,0.2,0.4$ from top
to bottom.
     } \label{fig3}
\end{figure}
 Actually, both of SBH and NBTZ have
the singe phase of the black hole because of their positive heat
capacities $C_{S/B}\ge 0$. A phase transition from the massless
black hole at $r_+=0(\rho_+=0)$ to the  black hole at
$r_+\not=0(\rho_+\not=0)$ is always possible to occur in the SBH
(NBTZ)~\cite{Myungbtz}. Now we study a phase transition between
two black hole for $r_+>0$. As is shown in Fig. 3, there is a
nonvanishing probability for decay of SBH into NBTZ  because of
Eq.(\ref{ief}). In order to express it more clearly, we consider
the difference between free energies. For $r_+>0$, the ground
state is the NBTZ. Hence we have to consider the free energy
difference  $\Delta F$ to show how a phase transition occurs
between them. We note that there is no critical temperature  for
$T>0$. Hence, the graph of $\Delta F$ shows that it is difficult
to identify the order of phase transition. This arises because
there exists a forbidden region of $r_+<0$.

\section{Transition between MTZ and TBH}

We wish  to study  the four-dimensional black hole with scalar
hair. For this purpose, we introduce the similar action~\cite{MTZ}
\beq I_4[g,\phi]= \int d^4x\sqrt{-g}\Bigg[\frac{R+6/l^2}{16 \pi
G_4}-\frac{1}{2}(\nabla\phi)^2-V(\phi)\Bigg] \label{4SDS} \eeq
where the potential $V(\phi)$ is given by \beq
V(\phi)=-\frac{3}{4\pi G_4l^2}\sinh^2 \Bigg[\sqrt{\frac{4 \pi
G_4}{3}}\phi\Bigg].\eeq Here $l$ is the curvature radius of
AdS$_4$ spacetimes. In the case of $\phi=0$, we have the
topological black hole solution with topology $\bf{R}^2 \times
\Sigma$, where $\Sigma$ is a two-dimensional hyperbolic manifold
of negative constant curvature ($k=-1$). The field equations are
\beqa
G_{\mu\nu}-\frac{3}{l^2}g_{\mu\nu}&=&8 \pi T_{\mu\nu},\\
\nabla^2\phi&=&V'(\phi) \eeqa with the stress-energy tensor
\begin{equation}
T_{\mu\nu}=\partial_\mu
\phi\partial_\nu\phi-\frac{1}{2}g_{\mu\nu}\Big(
\partial_\eta\phi\partial^\eta \phi-2V(\phi)\Big).
\end{equation}

Then the solution of the MTZ black hole is give by \beq
ds^2_{M}=\frac{r(r+2G_4\mu)}{(r+G_4\mu)^2}\Bigg[-f(r)dt^2+
\frac{dr^2}{f(r)}+r^2d\sigma^2\Bigg]\eeq where the metric function
$f(r)$ is given by  \beq
f(r)=\frac{r^2}{l^2}-\Bigg(1+\frac{G_4\mu}{r}\Bigg)^2 \eeq and the
scalar field has the configuration \beq
\bar{\phi}(r)=\sqrt{\frac{3}{4 \pi
G_4}}\tanh^{-1}\Big[\frac{G_4\mu}{r+G_4\mu}\Big]. \eeq Here $\mu$
is a parameter, playing the role of the reduced mass.
 The event (cosmological) horizons are  determined by solving $f(r)=0(r^2-l(r+G_4\mu)=0)$ as  \beq
\label{4EH} r_{\pm}= \frac{l}{2}\Bigg[1\pm
\sqrt{1+\frac{4G_4\mu}{l}}\Bigg]\eeq provided the reduced mass is
bounded from below by $\mu \ge -l/4G_4$. For $\mu=-l/4G_4$, we
have the degenerate horizon at $r_+=r_-=l/2$. As is shown in Fig.
4, we have the range of $0\le r_- \le l/2$ and $r_+\ge l/2$. Hence
the causal structure is not the Schwarzschild-AdS black hole
replacing $S^2$ by $\Sigma$~\cite{MTZ} but the BTZ black hole $S$
by $\Sigma$~\cite{BTZ}. For the massless case of
$\mu=0(\bar{\phi}=0)$, we have the metric \beq
ds^2=-\Big(\frac{r^2}{l^2}-1\Big)dt^2+\Big(\frac{r^2}{l^2}-1\Big)^{-1}dr^2
+r^2d\sigma^2 \eeq which is a locally AdS$_4$ spacetime but has a
topological horizon at $r_+=l$~\cite{BM}.

For the case of $\phi=0(V(\phi)=0)$, we have the vacuum solution
of topological black hole with hyperbolic horizon (TBH) \beq
ds^2_{T}=-\Big(\frac{\rho^2}{l^2}-1-\frac{2G_4
\mu_0}{\rho}\Big)dt^2+\Big(\frac{\rho^2}{l^2}-1-\frac{2G_4
\mu_0}{\rho}\Big)^{-1}d\rho^2 +\rho^2d\sigma^2. \eeq

\begin{figure}[t!]
 \centering
   \includegraphics{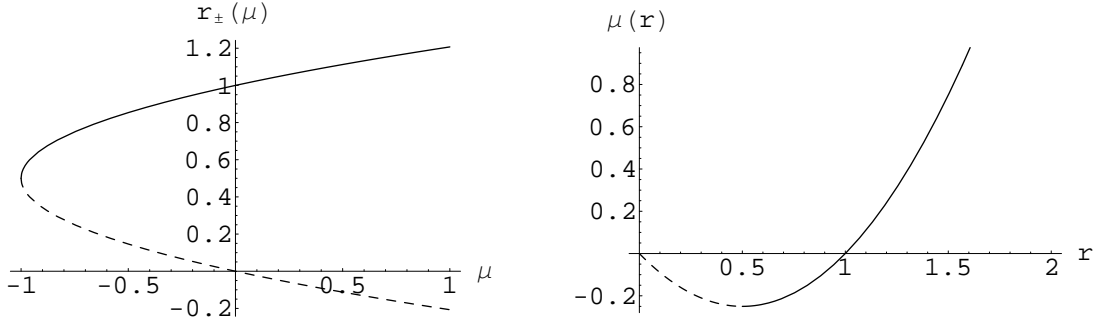}
    \caption{Left panel: plot of the event horizon $r_+$ (solid curve) and  the cosmological horizon $ r_-$ (dashed curve)
as  functions of reduced mass $\mu$.  Right panel: its inverse
function $\mu(r_\mp)$.
    $\mu(r_+)$ (solid curve) represents the mass $M_M$ of MTZ  for $\sigma/4\pi G_4=1$.
     Here we choose $G_4=l/4$ with $l=1$ for numerical computations.  } \label{fig4}
\end{figure}

 Thermodynamic quantities of the MTZ  are given by
 Hawking temperature $T_M$, mass $M_M$, heat capacity $C_M$, entropy
 $S_M$, and
 on-shell free energy $F_M$
by \beqa \label{MBH}T_M&=&\frac{1}{2\pi
l}\Big(\frac{2r_+}{l}-1\Big),~M_M=\frac{\sigma r_+}{4\pi
G_4}\Big(\frac{r_+}{l}-1\Big),\\
C_M&=&\frac{\sigma l}{4 G_4}\Big(2r_+-l\Big)=S_M,~
F_M(r_+)=-\frac{\sigma}{8 \pi G_4}\Big(\frac{2r_+^2}{l}-2r_+
+l\Big), \eeqa where $r_+ \ge l/2$ and $\sigma$ denotes the area
of $\Sigma$. We note that the first-law of thermodynamics
$dM_M=T_MdS_M$ holds for the MTZ. The free energy can be rewritten
as a function of $T_M$ as
 \beq F_M(T_M)=-\frac{\sigma l}{8 \pi G_4}\Bigg(1-2\pi (T_M+T_c)+2\pi^2 (T_M+T_c)^2\Bigg) \eeq
with $T_c=1/2\pi l$. Here we define the off-shell  free energy as
a function of $r_+$ and $T$ \beq F^{off}_M(r_+,T)=M_M-T S_M. \eeq
Here the horizon radius $r_+$ is the order parameter and the
external temperature $T$ is the control parameter for the
description of the phase transition. On the other hand, the TBH
provides thermodynamic quantities as Hawking temperature $T_T$,
mass $M_T$, heat capacity $C_T$, entropy $S_T$, and
 on-shell free energy $F_T$
 \beqa \label{TBH}
T_T&=&\frac{1}{4\pi
\rho_+}\Big(\frac{3\rho_+^2}{l^2}-1\Big),~M_T=\frac{\sigma
\rho_+}{8\pi G_4 }\Big(\frac{\rho_+^2}{l^2}-1\Big),\\
C_T&=& \frac{\sigma
\rho_+^2(3\rho_+^2-l^2)}{2G_4(3\rho_+^2+l^2)},~S_T=\frac{\sigma
\rho_+^2}{4 G_4 },~ F_T(\rho_+)=-\frac{\sigma \rho_+}{16 \pi G_4
}\Big(\frac{\rho_+^2}{l^2}+1\Big) \eeqa with $\rho_+ \ge
l/\sqrt{3}$. Solving the equation of $3\rho_+^3-4 \pi l^2 T_T
\rho_+-l^2=0$, one finds $\rho_+=\rho_+(T_T)$. Plugging this into
$F_T(\rho_+)$ leads to   \beq F_T(T_T)=-\frac{\sigma }{16 \pi
G_4}\Bigg[ \Bigg(\frac{2 \pi l^2 T_T+\sqrt{3l^2+( 2\pi
l^2T_T)^2}}{3l^{2/3}}\Bigg)^3+\frac{2 \pi l^2 T_T+\sqrt{3l^2+(2
\pi l^2T_T)^2}}{3}\Bigg]. \eeq
 Here we define the off-shell free
energy as a function of $\rho_+$ and $T$ \beq
F^{off}_T(\rho_+,T)=M_T-T S_T. \eeq
\begin{figure}[t!]
 \centering
   \includegraphics{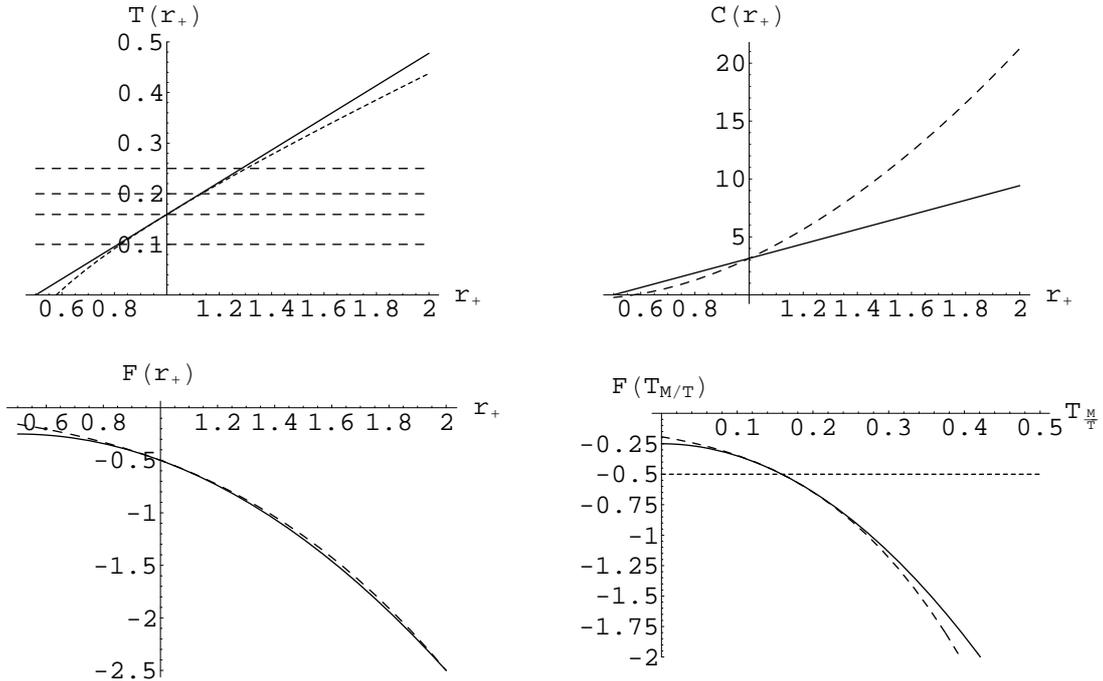}
    \caption{Coordinate matching of $\rho_+=r_+$ with $\sigma/4\pi G_4=1$.
     Hawking temperature $T(r_+)$  with external temperatures $T=0.25,0.2,0.16,0.1$
from top to bottom, heat capacity $C(r_+)$, free energy
    $F(r_+)$ and $F(T_{M/T})$  for MTZ (solid curves) and
    TBH (dashed curves).
     } \label{fig5}
\end{figure}
For  coordinate matching of $r_+=\rho_+$ with $l=1$,  we observe
from Fig. 5 that inconsistent equalities of free energy appear for
$r_+>1$ \beqa F_T(r_+)&>& F_M(r_+) ~~{\rm for}~~ r_+<1~~{\rm
and}~~ F_T(T)>
F_M(T)~~{\rm for}~~ T<T_c,\\
F_T(r_+)&>& F_M(r_+) ~~{\rm for}~~ r_+>1~~{\rm and}~~ F_T(T)<
F_M(T)~~{\rm for}~~ T>T_c.\eeqa  Hence we do not make a further
progress on  what kind of the phase transition happens for the
coordinate matching.

In order to resolve this situation, we need to introduce the
temperature matching: $T_S=T_B$, which means that
$r_+=\frac{3\rho_+}{4}- \frac{1}{4\rho_+}+\frac{1}{2}$ with
$\rho_+ \ge r_+$. Then we find from Fig. 6 that   consistent
equalities are found  for free energy \beqa F_T(\rho_+)&>&
F_M(\rho_+) ~~{\rm for}~~ \rho_+<1~~{\rm and}~~ F_T(T)>
F_M(T)~~{\rm for}~~ T<T_c,\\
F_T(\rho_+)&<& F_M(\rho_+) ~~{\rm for}~~ \rho_+>1~~{\rm and}~~
F_T(T)< F_M(T)~~{\rm for}~~ T>T_c.\eeqa
\begin{figure}[t!]
 \centering
   \includegraphics{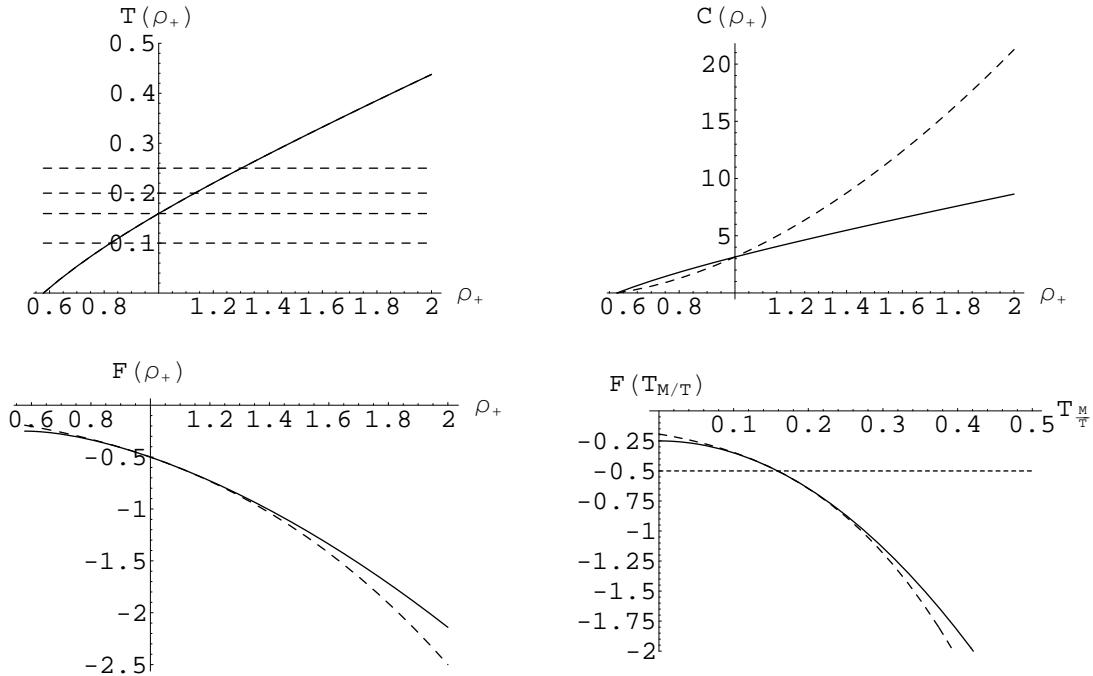}
    \caption{Temperature matching of $r_+=\frac{3\rho_+}{4}-
     \frac{1}{4\rho_+}+\frac{1}{2}$ with $\sigma/4\pi G_4=1$.
     Hawking temperature $T(\rho_+)$ with external temperatures $T=0.25,0.2,0.16,0.1$,
      heat capacity $C(\rho_+)$, free energy
    $F(\rho_+)$ and $F(T_{M/T})$  for MTZ (solid curves) and
    TBH (dashed  curves).
     } \label{fig6}
\end{figure}
As is shown in Fig. 6, we could  find the critical temperature
$T_c=1/2\pi=0.16$ where $F_M(T_c)=F_T(T_c)$ at $\rho_+=1$. Hence,
we separate the whole region into  the left region of
$1/\sqrt{3}\le \rho_+ <1$ and the right region of $\rho_+>1$ for
the temperature matching.
 For
$\rho_+<1$ the MTZ configuration is more favorable than the TBH,
while for $\rho_+>1$, the TBH configuration is more favorable than
the MTZ. This means that for $\rho_+ <1$, the ground state is the
MTZ, whereas for $\rho_+>1$, the ground state is chosen to be the
TBH.
 Actually, both of MTZ and TBH have
the single phase of the black hole because of their positive heat
capacities $C_{S/B}\ge 0$. Also we note that the heat capacities
are nothing special at the critical point $\rho_+=1$. Hence phase
transitions from the extremal black hole at
$\rho_+=1/\sqrt{3}(r_+=1/2)$ to the non-extremal black hole at
$\rho_+>1/\sqrt{3}(r_+>1/2)$ are always possible to occur in the
TBH (MTZ), respectively.

Now we are in a position to  identify what kind of the phase
transitions is going on for the temperature matching. We remind
the reader that in this work, $r_+(\rho_+)$ are the order
parameters and $T$ is the control parameter for the description of
the phase transition in the MTZ-TBH system~\footnote{We do not use
$\lambda_{\phi}=|\tanh\sqrt{4\pi
G_4/3}\bar{\phi}(r_+)|=|(r_+-1)/r_+|$ as the order parameter for
$T<T_c$ ~\cite{MTZ,KMPS} because it vanishes for $T>T_c$.}.

 It was
conjectured that this  is  a second-order phase transition as
occurred in the ferromagnetic system.
\begin{figure}[t!]
 \centering
   \includegraphics{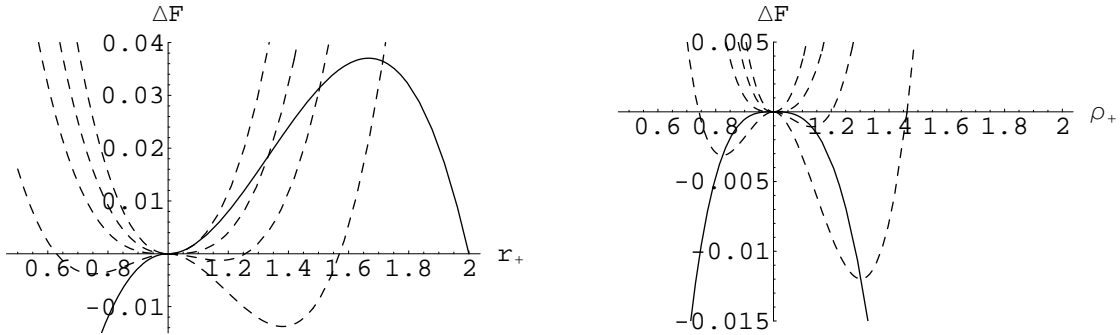}
    \caption{
The left panel of coordinate matching shows the difference of the
on-shell (solid curve)  free
    energies  $\Delta F_{L/R}(r_+)$ and that of off-shell (dashed curve) free
energies of $\Delta F^{off}_{L/R}(r_+,T)$. The right panel denotes
$\Delta F_{L/R}(\rho_+)$ and  $\Delta F^{off}_{L/R}(\rho_+,T)$ for
the temperature matching. Here $T=0.25,0.2,0.16,0.1$ for
$r_+(\rho_+)<1$ and  $T=0.1,0.16,0.2,0.25$ for $r_+(\rho_+)>1$
from top to bottom. } \label{fig7}
\end{figure}
In order to prove it,  we define the difference between on-shell
(off-shell) free energies on both sides \beqa \Delta
F_{L/R}(r)&=&\pm\Big[F_M(r)-
F_T(r)\Big],\\
 \Delta F^{off}_{L/R}(r,T)&=&\pm\Big[F^{off}_M(r,T)-F^{off}_T(r,T)\Big]\eeqa
 with $r=r_+(\rho_+)$ for coordinate (temperature) matchings, respectively.
 These are depicted in Fig. 7.
For the coordinate matching, there is no saddle point of thermal
equilibrium as the crossing point between $\Delta F_{L/R}(r_+)$
and $\Delta F^{off}_{L/R}(r_+,T)$. However, for the temperature
matching, there exist  saddle points of thermal equilibria as the
crossing points between $\Delta F_{L/R}(\rho_+)$ and $\Delta
F^{off}_{L/R}(\rho_+,T)$.

Firstly, we consider the phase transition between two black hole
on the left region of $1/\sqrt{3}\le \rho_+ <1$ . As is shown in
Fig. 6, there is a nonvanishing probability for decay of TBH into
MTZ. In order to express it more precisely, we have to consider
the difference between free energies.  The graph of $\Delta
F^{off}_{L}$ in Fig. 7 shows that there is a transition from TBH
to MTZ configuration. For $T=0.1<T_c$, we find that the dominance
of system is the MTZ, while for $T=0.2,0.25>T_c$ the dominance is
changed to be  the TBH.

Secondly,  we consider the phase transition between two black hole
on the right region of $ \rho_+ >1$.  As is shown in Fig. 6, there
is a nonvanishing probability for decay of MTZ into TBH. The graph
of $\Delta F^{off}_{R}$ in Fig. 7 indicates that there is a
transition from MTZ to TBH configuration. There is a change of
dominance at the critical temperature $T=T_c$. For $T=0.1<T_c$, we
find that the dominance of system is the MTZ, while for
$T=0.2,0.25>T_c$ the dominance is given by the TBH. As the whole
picture, for $T<T_c$ the MTZ is ground state while for $T>T_c$,
the TBH is the ground state. However, as $T\to 0$, the MTZ
configuration may lead to the extremal MTZ with $T_M=0$ at
$\rho_+=1/\sqrt{3}$ by evaporating process. Hence the non-extremal
MTZ may not be   a truly ground state near $T=0$.

A symmetric configuration of $\Delta F^{off}_{L/R}$ around the
critical point $\rho_+=1$ is expected  to exist for a typical
feature of the second-order phase transition in condensed matter
physics~\cite{Kol}. Even though it is asymmetric in the MTZ-TBH
system of black hole  physics, it shows still the nature of
second-order phase transition. To compare this with the
first-order phase transition, one simply refers the Hawking-Page
phase transition in the Schwarzschild-AdS black
hole~\cite{myungsds,CM}.

\section{Quasinormal modes and phase transitions}
Recently, black holes' quasinormal modes (QNMs) can reflect the
black hole phase transition. It was claimed that the behavior of
QNMs around $r_+=1(\rho_+=1)$  shows a signal for the second-order
phase transition by calculating the electromagnetic perturbations
on the background of the MTZ  and TBH~\cite{KMPS}. Furthermore,
there was a change in the slope in the $w_R$-$w_I$ diagrams:
positive for $r_+=0.97$, infinity at $r_+=1$, and negative for
$r_+=1.03$ by calculating the scalar perturbations on the
background of the MTZ~\cite{SWSLC}. Also the same thing happened:
positive for $\rho_+=0.97$, infinity at $\rho_+=1$, and negative
for $\rho_+=1.03$  on  the background of the TBH. So it is very
important to justify whether the QNMs plays the role of  an
effective tool to disclose a phase transition in the general black
hole background. However, there is no change in the slope on the
$k=0$ AdS black hole background which was introduced to explain
the phase transition between the $k=0$ AdS black hole and AdS
soliton. Of course, there is no QNMs on the AdS soliton background
because of the absence of event horizon.

Without the temperature matching, the points of $r_+=1$ in the MTZ
and $\rho_+=1$ in the TBH are nothing special thermodynamically.
That is, their positive heat capacities are continuous across at
these points. These are not the point of the minimum temperature
in the Schwarzschild-AdS black hole~\cite{myungsds} and Davies'
point in the Reissner-Nordstr\"om black hole~\cite{davies}, where
heat capacities blow up~\cite{CEJM,RWY,myungrn}. However, as is
shown Fig. 5, two black holes have the same thermodynamic
quantities at $r_+=\rho_+=1$.  An interesting point is that two
black holes have the same free energy as
$F_M(r_+=1)=F_T(\rho_+=1)=-0.5~(F_M(T_M=T_c)=F_T(T_T=T_c)=-0.5)$.
Also this free energy corresponds to that of topological black
hole with zero mass. It is turned out that this black hole is
stable~\cite{BM}, which can be confirmed by the positive heat
capacity. Hence $r_+=1(\rho_+=1)$ have  noting to do with the
thermodynamic instability.

As was explained in the previous section, the second-order phase
transition between MTZ and TBH at $\rho_+=1(T=T_c)$ is shown to
occur by using temperature matching between MTZ and TBH and the
difference of their free energies. Even though the QNMs may give a
signal to the phase transition, this is not a deterministic
evidence for the second-order phase transition between MTZ and
TBH.

\section{Discussions}
We start with discussion on three-dimensional black holes. We note
that for three dimensions, the $k=0,\pm1$ AdS$_3$ black holes are
the same and AdS$_3$ soliton is diffeomorphic to thermal AdS$_3$
spacetimes~\cite{SSW}. Hence we could not distinguish topological
black holes in three dimensions. Introducing a single
self-interacting scalar field minimally coupled to gravity, then
one finds the SBH.  However, there is no non-zero critical
temperature of $T_S=T_B$ for coordinate matching.  Hence the
transition between massless black hole at $r_+=0(\rho_+=0)$ and
SBH(NBTZ) is possible to occur~\cite{Myungbtz}. Two temperatures
become the same only when $T_S=T_B=0$. This point contrasts to the
four-dimensional critical temperature with $T=T_c=1/2\pi l$.

Requiring the temperature matching, there is a nonvanishing
probability for decay of the SBH into the NBTZ  for $r_+>0$.
However, we could not identify its order because there is no
critical temperature. Concerning the QNMs, we have the results:
AdS soliton (thermal AdS) and massless black hole at $r_+=0$ $\to$
No QNMs and NBTZ  with $r_+ \not= 0\to$ QNMs~\cite{ML}. We expect
that the same thing happens for the SBH: massless black hole at
$\rho_+=0$ $\to$ No QNMs and scalar black hole with $\rho_+ \not=
0$ $\to$ QNMs. This implies that the QNMs may not show a direct
evidence for the phase transition between SBH and NBTZ.

\begin{table}
\centering
\begin{tabular}{|c||c|c|c|}
\hline
 &starting configuration&  ending configuration & remarks \\
\hline
    $k=1$   & thermal AdS$_4$ space & large black hole & first-order (HPT) \\ \hline
  $k=0$ & AdS$_4$ soliton & large black hole & ?(HPT) \\ \hline
  $k=-1$ & EBH & TBH & ?(HPT)\\
  \hline
  $k=-1$ & EBH &  MTZ &
  ?(HPT)\\ \hline \hline
  $k=-1$ & MTZ  &  TBH & second-order
  \\ \hline
\end{tabular} \caption{Summary of phase transitions  in asymptotically AdS$_4$ spacetimes.
EBH represents the extremal black hole with degenerate horizon and
 HPT denotes the Hawking-Page phase transition.}
\end{table}
We summarize all possible phase transitions in AdS$_4$ spacetimes
in Table 1. First of all, for $k=1$ spherical horizon, we have the
Hawking-Page phase transition between thermal AdS$_4$ and large
black hole in the Schwarzschild-AdS black hole. This  is the
first-order phase transition. In the case of $k=0$ Ricci-flat
horizon, it is conjectured that there may exist a phase transition
between AdS$_4$ soliton and large black hole with Ricci-flat
horizon. This is similar to the Hawking-Page phase transition but
there is no unstable small black hole as a mediator of the phase
transition. Hence it is unclear to define its order of phase
transition. For the case of $k=-1$ hyperbolic horizon, we expect
that there exists a phase transition between extremal and
non-extremal black holes. This happens for MTZ and TBH,
respectively. The phase transition is a kind of Hawking-Page
transition and its  order  is not clearly determined.

Finally, we observe a second-order phase transition between MTZ
 and TBH at the critical temperature $T=T_c$.

In conclusion, we explicitly show that the phase transition
between MTZ  and TBH is second-order by using the temperature
matching and difference of free energies. Without temperature
matching, we do not confirm that the phase transition is
second-order.

\section*{Acknowledgments}
This work was supported by the Korea Research Foundation
(KRF-2006-311-C00249) funded by the Korea Government (MOEHRD).

\end{document}